\documentclass[9pt,twocolumn,twoside]{optica}
\setboolean{shortarticle}{true}
\setboolean{minireview}{false}
\title{A Chip-scale, Full-Stokes Polarimeter}

\usepackage[numbers]{natbib}

\author{Zhongjin Lin}
\author{Leslie Rusch}
\author{Yuxuan Chen}
\author[*]{Wei Shi}

\affil{Department of Electrical and Computer Engineering, Center for Optics, Photonics and Lasers (COPL), Universit\'e Laval, Qu\'ebec, QC, Canada G1V 0A6}

\affil[*]{Corresponding author: wei.shi@gel.ulaval.ca}

\ociscodes{(230.3120) Integrated optics devices; (120.5410)   Polarimetry; (130.6622) Subsystem integration and techniques.}

\begin{abstract}

The polarization of light conveys unique information that can be exploited by crucial applications. The bulky and costly discrete optical components used in conventional polarimeters limit their broad adoption.
A compact, low-cost polarimeter would bring this functionality into a myriad of new scenarios and revolutionize its exploitation. Here we present a high-performance, full-Stokes polarimeter on a silicon chip. A surface polarization splitter and  on-chip optical interferometer circuit produce the analysis matrix of an optimally conditioned polarimeter. This solid-state polarimeter is a system-on-a-chip with exceptional compactness, stability, and speed that could be used singly or in integrated arrays. Large arrays can increase the speed and resolution of full-Stokes imaging; therefore, our design provides a scalable polarimeter solution.

\end{abstract}

\setboolean{displaycopyright}{true}

\begin{document}
\maketitle
Characterization of the state of polarization (SoP) of light is crucial for many important applications.
In the field of astronomy, polarimetry can characterize the rotation of stars \cite{cotton2017polarization} and their stellar magnetic fields\cite{chrysostomou2007circular}, which cannot be accurately measured using other properties of light.
Polarimetry is also a powerful tool to characterize aerosol particles, a  subject of increasing importance to both human health and our natural environment \cite{zhang2015fine}.
Although the earliest polarimeter dates back to the 1850s,\cite{stokes1852} state-of-the-art commercial polarimeters  remain bulky, or unable to provide all four elements of the Stokes tensor, i.e., full-Stokes functionality.
Recently, some miniaturized full-Stokes polarimeters have been proposed.\cite{wu2017visible,mueller2016ultracompact,espinosa2017chip,afshinmanesh2012measurement} Capasso and co-workers designed an ultra-compact in-line polarimeter using polarization-dependent scattered fields of plasmonic metasurface.\cite{mueller2016ultracompact} Their polarimeter requires an off-chip camera to collect the scatted light, and their use of metallic nanoantennas creates parasitic absorption losses.

\begin{figure} [h!]
\centering
\includegraphics[width=70mm]{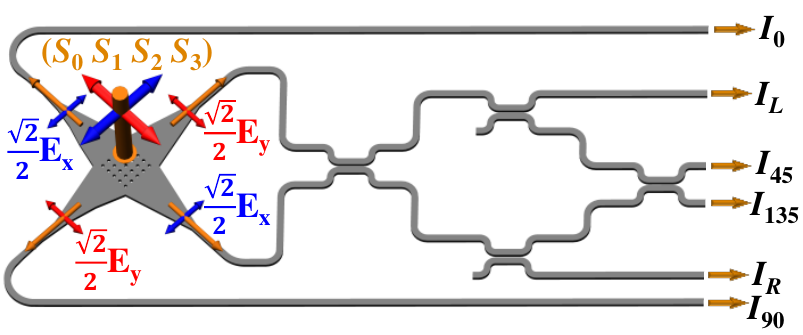}
\caption{A schematic of the proposed polarimeter. Incident light is decomposed into two orthogonal linear-polarization components (indicated by the blue and red arrows) that are split and coupled to four waveguides. The orange arrows point to the propagation direction of light. The outputs of the six ports are connected to photodetectors for intensity readouts: $I_0, I_{L}, I_{45}, I_{135}, I_{R}$, and $I_{90}$. Four directional couplers compose an optical interferometer circuit to calculate the third and fourth Stokes parameters.}
\label{schematic_of_device}
\end{figure}

\begin{figure*}[ht]
\centering\includegraphics[width=140 mm]{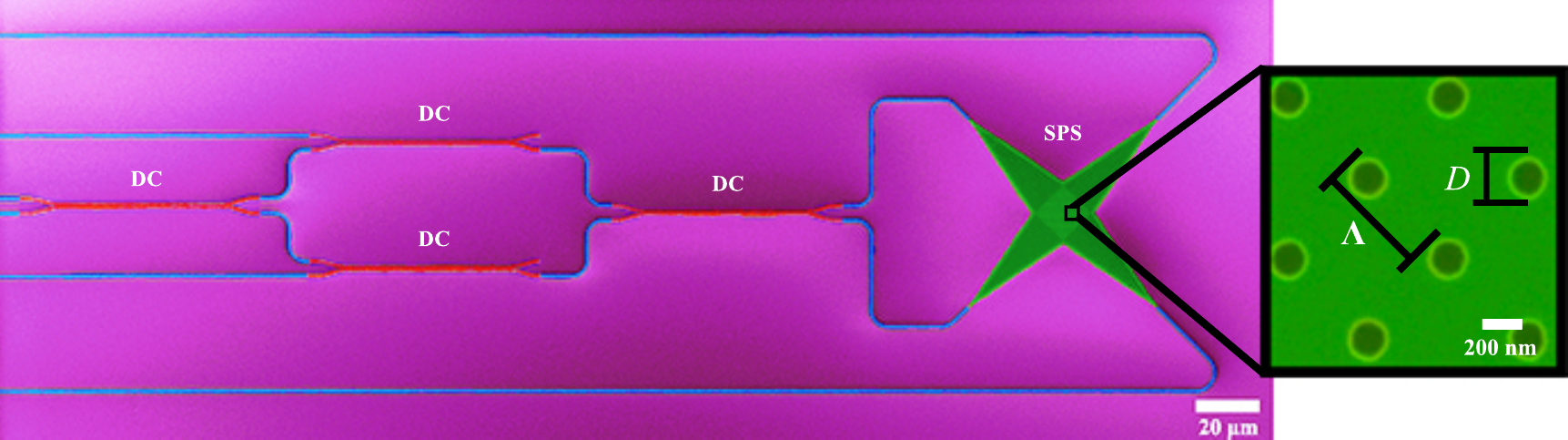}
\caption{SEM images of the fabricated device. The green, blue, and red areas are the surface polarization splitter (SPS), waveguides, and directional couplers, respectively. The inset shows an enlarged image of the SPS.}
\label{SEM}
\end{figure*}

Silicon photonics brings to the manipulation light the economies of scale and advantages of tremendous integration long enjoyed by VLSI in electronics. Silicon photonics can integrate a vast number of optical components on a single chip in proximity with microelectronics. The manipulation of light is therefore enhanced with digital signal processing to create complete systems-on-a-chip. While silicon photonics initially targeted optical interconnects, a broader range of applications is now under development, such as sensors, light detection and ranging (LiDAR), and imaging.\cite{Simplysilicon}  Large-scale silicon photonic integrated circuits have been demonstrated\cite{alfano2016optical,sun2013large,shen2015integrated,cai2012integrated,lu2015broadband,luo2014wdm},
and are being turned to the polarimetry application. Mart\'{i}nez, \emph{et al.}, used the spin-orbit interaction of light to demonstrate the use of sub-wavelength scattering on silicon for local observation of SoP.\cite{espinosa2017chip} They used the careful manipulation of a metallic nanoparticle or supplemental polarization filtering for their SoP characterization, however, the accuracy of their polarimeter has yet to be established with either method. A silicon photonic Stokes vector receiver has been demonstrated for high-speed optical communications.\cite{dong2016128} It uses a nanotaper on the edge of a chip to collect light from a fiber;  on-chip components such as beam splitters,  polarization rotators, and optical hybrids are used for polarization decomposition and phase readouts. This structure, however, cannot be used to realize two dimensional arrays for imaging polarimetry.

We present a novel chip-scale, full-Stokes polarimeter in silicon photonics with proven SoP accuracy, compatibility with imaging polarimetry, and scalability to large arrays to enhance speed and resolution. Our polarimeter consists of a surface polarization splitter (SPS) and on-chip optical interferometers that convert the SoP directly to intensity readouts. In addition to the fiber-optic application of sensor and communication, our compact polarimeter element can be arrayed for free-space applications such as polarimetric LiDAR and imaging polarimetry. The proposed chip-scale polarimeter is based on the standard 220-nm-thick silicon-on-insulator (SOI) wafer with a 2~$\mu$m buried oxide layer and 2-$\mu$m oxide cladding.

The schematic of the proposed device is shown in Fig.~\ref{schematic_of_device}. A two-dimensional dielectric grating structure \cite{taillaert2003compact} is used as a SPS to decompose incoming light with an arbitrary SoP, $(S_0,S_1,S_2,S_3)$, into two orthogonal linearly polarized E-field components, $\bf E_x$ and $\bf E_y$. The E-field components are separated; each has it power split (ideally a 50:50 split) and coupled into two single-mode waveguides that exit the structure in opposite directions. Note that the excited optical waves propagating in the four waveguides, $\frac{\sqrt{2}}{2}\mathbf{E_x}, \frac{\sqrt{2}}{2}\mathbf{E_y}, \frac{\sqrt{2}}{2}\mathbf{E_x}, \frac{\sqrt{2}}{2}\mathbf{E_y}$, carry full SoP information on the incoming light. The SoP can be retrieved using their intensities and the relative phase between the two orthogonal E-field components. Since most photo-detectors (PDs) are only sensitive to intensity, an optical interferometer circuit is designed to convert phase information into intensity. The interferometer is realized via four directional couplers (DCs).

As shown in Fig.\ref{schematic_of_device}, the device has six outputs.
Four outputs provide intensity measurements with information on the linear polarization of incident light at specific directional rotations: linear horizontal ($I_0$), linear $ 45^{\circ} $ ($I_{45}$),  linear vertical ($I_{45}$), linear $ 135^{\circ} $($I_{135} $).
Two outputs provide intensity measurements with information on the circular polarization: right-handed circular ($I_R$), and left-handed circular ($I_L$) polarization.

Straight paths without directional couplers provide 
\begin{equation}
I_0\propto \frac{1}{2}\mid \mathbf{E_x} \mid ^2
\label{eq1}
\end{equation}
\begin{equation}
I_{90}\propto \frac{1}{2}\mid \mathbf{E_y} \mid ^2
\label{eq2}
\end{equation}
The interferometric structure provides
\begin{equation}
I_L\propto \frac{1}{8}\mid \mathbf{E_x}e^{-i\frac{\pi}{2}}+\mathbf{E_y} \mid ^2
\label{eq3}
\end{equation}
\begin{equation}
I_{45}\propto \frac{1}{8}\mid \mathbf{E_x}+\mathbf{E_y} \mid ^2
\label{eq4}
\end{equation}
\begin{equation}
I_{135}\propto \frac{1}{8}\mid \mathbf{E_x}-\mathbf{E_y} \mid ^2
\label{eq5}
\end{equation}
\begin{equation}
I_R\propto \frac{1}{8}\mid \mathbf{E_x}+\mathbf{E_y}e^{-i\frac{\pi}{2}} \mid ^2
\label{eq6}
\end{equation}

\begin{figure} [h!]
\centering\includegraphics[width=70 mm]{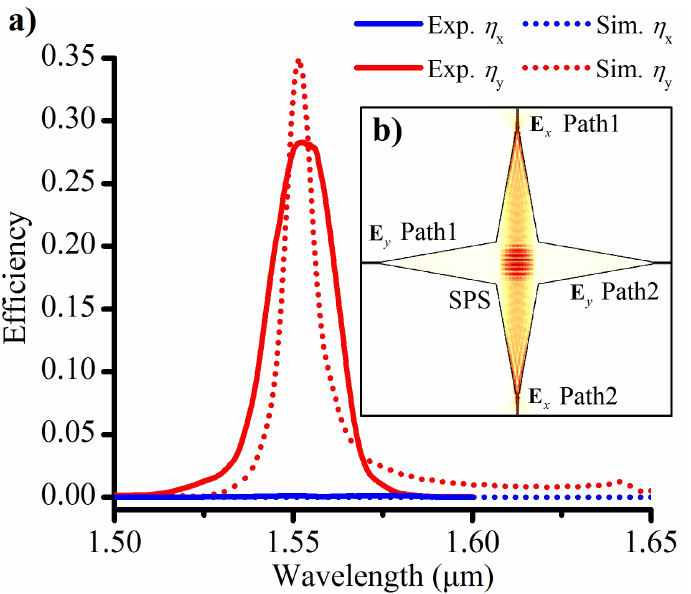}
\caption{Performance of the SPS. For input of $x$-polarized light at normal incidence, \textbf{a)} measured (solid) and simulated (dotted) coupling efficiency of light propagating in vertical waveguide (red), \emph{i.e.}, $y$-direction ($\eta_y$), and horizontal waveguide (blue),\emph{ i.e.}, $x$-direction ($\eta_x$); \textbf{b)} The simulated intensity distributions within the SPS under the $x$-polarized light at 1550~nm wavelength.}
\label{SPS}
\end{figure}

\begin{figure*}[ht]
\centering\includegraphics[width=145 mm]{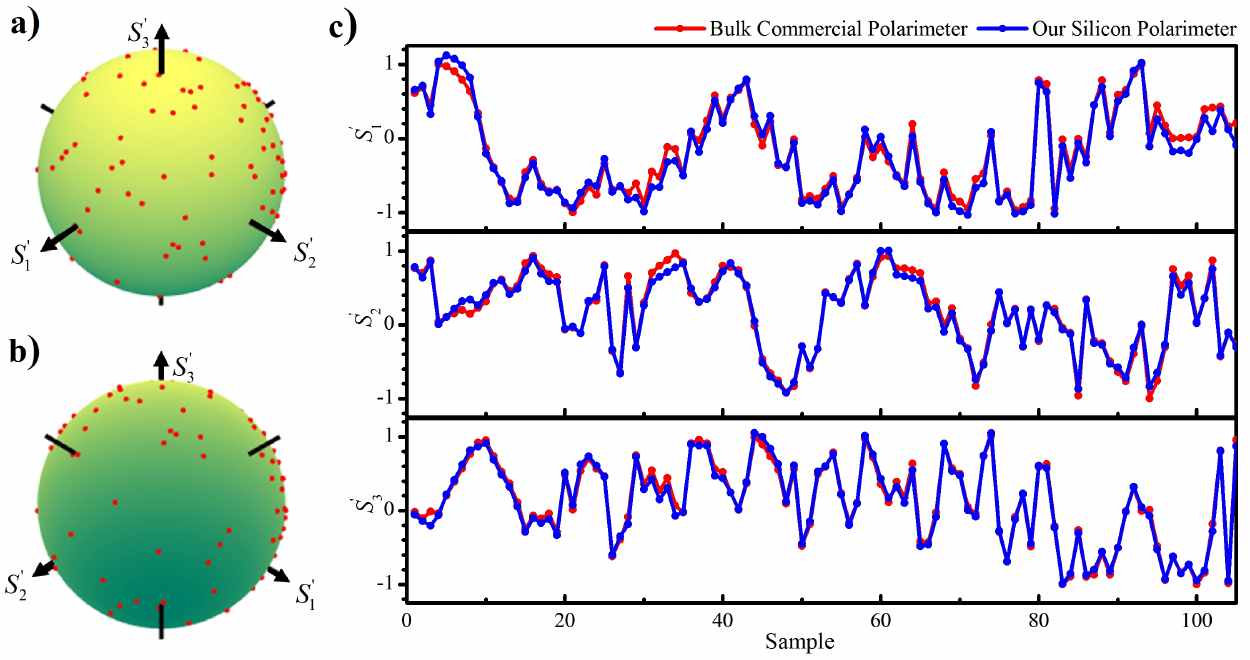}
\caption{Experimental results. \textbf{a)} front and \textbf{b)} back views of Poincar\'{e} sphere with red dots indicating incident polarization states dispersed randomly;  \textbf{c)} measured normalized Stokes vector $\mathbf{S'}=\left( S_0, S_1, S_2, S_3\right) ^T/S_0$ using our device (blue) and a commercial in-line polarimeter (red).}
\label{Experiment}
\end{figure*}
We write the full-Stokes vector as $\mathbf{S}=\left( S_0, S_1, S_2, S_3\right) ^T$ and the intensity vector as $\mathbf{I}= \left( I_0/2, I_{L}, I_{45}, I_{135}, I_{R}, I_{90}/2 \right) ^T$. The full-Stokes vector can be recovered from the six outputs via
\begin{equation}
\mathbf{S}\propto \mathbf{M_S}\cdot \mathbf{I}
\label{eq9}
\end{equation}
where $\mathbf{M_S} $ is the synthesis matrix of the polarimeter, given by
\begin{equation}
 \mathbf{M_S}=4*\left( \begin{array}{cccccc}
1 & 0 & 0 & 0 & 0 & 1 \\ 1 & 0 & 0 & 0 & 0 & -1  \\ 0 & 0 & 1 & -1 & 0 & 0 \\ 0 & 1 & 0 & 0 & -1 & 0
\end{array}  \right)
\label{eq10}
\end{equation}

Equation~\ref{eq10} is only valid under ideal conditions.   Imperfections such as imbalanced splitting ratios of the SPS and DCs, waveguides losses, and phase errors should be taken into consideration. For a general expression, we rewrite the E-fields in the four waveguides coupled from the SPS as $\kappa _{x1}\mathbf{E_x}e^{i\delta _{x1}}$,$\kappa _{x2}\mathbf{E_x}e^{i\delta _{x2}}$, $\kappa _{y1}\mathbf{E_y}e^{i\delta _{y1}}$, and $\kappa _{y2}\mathbf{E_y}e^{i\delta _{y2}}$.
The $\kappa$ coefficients represent the impact of imbalanced splitting ratio of the SPS and waveguide losses; the $\delta$ coefficients represent the impact of phase errors. The first index refers to the E-field component ($x$ or $y$), and the second index to one of the two output waveguides.
The straight-through and cross-coupling
coefficients of the DC are represented by $\tau _{DC}$ and $\kappa _{DC}$, respectively.
A calibrated synthesis matrix $\mathbf{M'_S}$ accounting for imperfections is given by
\begin{equation}
\mathbf{M'_S}=2*\left( \begin{array}{cccccc}
\frac{2}{\kappa _{x1}^2} & 0 & 0 & 0 & 0  & \frac{2}{\kappa _{y1}^2} \\ \frac{2}{\kappa _{x1}^2} & 0 & 0 & 0 & 0 & -\frac{2}{\kappa _{y1}^2} \\ 0  & -a  & \frac{b}{c} & \frac{-b}{c} & a &  \frac{2ad\kappa ^2 _{y2}}{\kappa ^2 _{y1}}\\ 0  & b  & \frac{a}{c} & \frac{-a}{c}& -b & \frac{2bd\kappa ^2 _{y2}}{\kappa ^2 _{y1}}
\end{array}  \right)
\label{eq11}
\end{equation}
where $a=\frac{\sin \left( \delta _{y2}-\delta _{x2}\right)}{\tau _{DC}\kappa _{DC}\kappa _{x2}\kappa _{y2}} $, $b=\frac{\cos \left( \delta _{y2}-\delta _{x2}\right)}{\tau _{DC}\kappa _{DC}\kappa _{x2}\kappa _{y2}} $, $c=2\tau _{DC}\kappa _{DC}$, and $d=\left(\kappa ^2 _{DC}-\tau ^2 _{DC}\right)$. More details about the theoretical analysis are given in Supplementary 1. In practice, the calibrated polarimetric matrix  $\mathbf{M'_S}$ can be obtained by four measurements of output intensities when inputting light with known, independent SoPs.

The device was designed and fabricated using a CMOS-compatible process with electron-beam lithography.
A scanning electron micrograph (SEM) of the fabricated device is presented in Fig.~\ref{SEM}. The SPS is formed using a $20\times20$ array of cylindrical holes fully etched through silicon with a period of $\Lambda=596$~nm and a hole diameter of $D=200$~nm (as shown in the inset of Fig.~\ref{SEM}). This design focused on the telecommunications wavelength band around 1550~nm, but could be directed to the band of interest for a given application. More details about the design of the SPS are given in Supplementary 2. The experimental and simulation results are shown in Fig.~\ref{SPS} when inputting (via normal incidence) $x$-polarized light. $\eta_x$ and $\eta_y$ are the efficiency of light coupled into x-directional and y-directional waveguides, respectively. The measured spectrum shows an efficiency ($\eta_y$) of near $27.7~\% $ at the wavelength of 1550~nm. A high extinction ratio of 35~dB was measured experimentally. An image of the simulated intensity distribution withing the SPS under the $x$-polarized light at 1550~nm wavelength is presented in Fig.~\ref{SPS}b. We observe strong coupling along the vertical direction where exit paths for $x$-polarized component of light are located. virtually no light is found in the horizontal direction where exit paths for $y$-polarized component of light are located. This is confirmed in experiment by the high extinction ratio.

Four known independent SoPs were used to calibrate the device and to calculate the system polarimetric matrix $\mathbf{M'_S}$ (Eq.~\ref{eq11}).
Then the performance of the polarimeter was experimentally tested using a series of SoPs  spread widely over the surface of the Poincar\'{e} sphere, as illustrated in Fig.~\ref{Experiment}a and b.
For each point in this series of randomly generated SoPs, we  simultaneously measured the SoP with our device and a commercial in-line polarimeter (details about the experiment are given in Supplementary 3).
At each measurement we normalized the SoP to a unitary first component, and plotted the remaining three components, \emph{i.e.},
$\mathbf{S'}=\left( S_0, S_1, S_2, S_3\right) ^T/S_0$.  The measured SoP results are summarized in Fig.~\ref{Experiment}c.
Excellent agreement is observed between the SoP measurements using our device and a commercial bench-top polarimeter. The root-mean-square error between measurements with the integrated and the bench-top instrument is $0.07$. This error is dominated by  intensity measurement errors due to the set-up variations and PD noise (Supplementary 5).\cite{tyo2002design} This error can be significantly reduced by PDs integrated on the same chip.
Notice that high-responsivity, high-speed Ge-on-Si PDs have already been demonstrated \cite{zhang2014high,derose2011ultra} and are now widely available in silicon photonics foundry processes.

\begin{figure}[h!]
\centering\includegraphics[width=60mm]{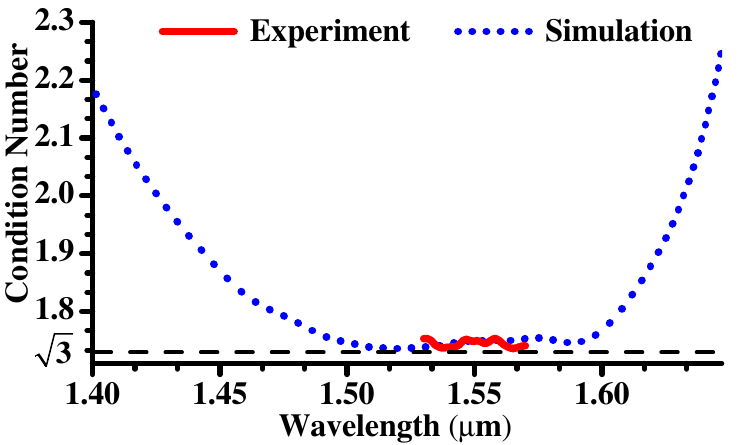}
\caption{The experimental and simulated  condition number of the device. The red and blue line are the experimental and numerical condition number $\kappa $ of the matrix $\mathbf{M'_S}$ as a function of the wavelength.}
\label{condition number}
\end{figure}

The condition number of the analysis matrix indicates how sensitive the reconstructed SOP is to  systematic errors such as miscalibration. The signal-to-noise ratio (SNR) of a polarimeter is determined by the condition number, with SNR maximized when the condition number is minimized.\cite{tyo2002design} The ideal polarimeter has a condition number of $\sqrt{3}$, the minimum value for full-Stokes analysis.\cite{tyo2002design} In addition to imperfections captured in (Eq.~\ref{eq11}), $\mathbf{M'_S}$ and its condition number is a function of wavelength due to the wavelength dependencies of the DCs and the SPS. To mitigate this effect, we used a broadband DC design with an asymmetric-waveguide-assisted section leading to a near 200~nm 1~dB bandwidth of splitting ratio (Supplementary Fig.~S8).\cite{lu2015broadband} The condition number as a function of wavelength is numerically simulated and is shown in Fig.~\ref{condition number}. We can observe that the curve shows a flat bottom very close to the optimal value $\sqrt{3}$ across a wide spectral range from $1.5$ to $1.6~\mu$m. The condition number is also calculated using the parameters (the coupling coefficients of the DC and the extinction ratio of the SPS) extracted from measurement (Supplementary 4), and agrees well with the numerical simulation (Fig. \ref{condition number}).

The demonstration of a silicon full-Stokes polarimeter paves the way to polarimetry sensor systems on a chip for a vast number of applications. Avoiding the use of free-space optical and mechanical components, this solid-state solution enables significant improvement in system robustness, size and cost. A polarimeter array can also be fabricated on a single chip with minimum increase in footprint and cost, proving a scalable solution for applications such as imaging polarimetry and polarimetric LiDAR. For large-scale arrays (e.g., in 2D polarization imaging), we can spatially separate the SPS and the optical interferometer circuit, which allows us to group the SPS elements in a compact footprint to achieve a large fill factor.
Furthermore, a number of the SPS elements can share one set of optical interferometer circuit and integrated PDs through on-chip optical switches \cite{seok2016large,almeida2004all} so that the SoP received by each SPS can be analyzed in a time series. Our device can also be used as a polarization analyzer for polarimetric fiber-optic sensors whose application is limited by the high cost of currently available polarimeters. The proposed structure can be applied to other CMOS-compatible materials such as silicon nitride and germanium for a broad spectrum from visible to mid-infrared.\cite{moss2013new,soref2010mid} Furthermore, it can be readily integrated with other silicon photonic functions such as spectrometers \cite{souza2018fourier} for a multi-dimensional optical measurement system on a chip.






\begin{thebibliography}{10}
\newcommand{\enquote}[1]{``#1''}

\bibitem{cotton2017polarization}
D.~V. Cotton, J.~Bailey, I.~D. Howarth, K.~Bott, L.~Kedziora-Chudczer,
  P.~Lucas, and J.~Hough, {\protect\JournalTitle{Nature Astronomy}} \textbf{1},
  690 (2017).

\bibitem{chrysostomou2007circular}
A.~Chrysostomou, P.~W. Lucas, and J.~H. Hough, {\protect\JournalTitle{Nature}}
  \textbf{450}, 71 (2007).

\bibitem{zhang2015fine}
Y.~Zhang and F.~Cao, \enquote{Fine particulate matter (pm2. 5) in china at a
  city level, sci. rep., 5, 14884,}  (2015).

\bibitem{stokes1852}
G.~G. Stokes \emph{et~al.}, {\protect\JournalTitle{Proc. Cambridge
  Philosophica}} \textbf{1}, 115 (1852).

\bibitem{wu2017visible}
P.~C. Wu, J.-W. Chen, C.-W. Yin, Y.-C. Lai, T.~L. Chung, C.~Y. Liao, B.~H.
  Chen, K.-W. Lee, C.-J. Chuang, C.-M. Wang \emph{et~al.},
  {\protect\JournalTitle{ACS Photonics}}  (2017).

\bibitem{mueller2016ultracompact}
J.~B. Mueller, K.~Leosson, and F.~Capasso, {\protect\JournalTitle{Optica}}
  \textbf{3}, 42 (2016).

\bibitem{espinosa2017chip}
A.~Espinosa-Soria, F.~J. Rodr{\'\i}guez-Fortu{\~n}o, A.~Griol, and
  A.~Mart{\'\i}nez, {\protect\JournalTitle{Nano letters}} \textbf{17}, 3139
  (2017).

\bibitem{afshinmanesh2012measurement}
F.~Afshinmanesh, J.~S. White, W.~Cai, and M.~L. Brongersma,
  {\protect\JournalTitle{Nanophotonics}} \textbf{1}, 125 (2012).

\bibitem{Simplysilicon}
Editor, {\protect\JournalTitle{Nature Photonics}} \textbf{4}, 491 (2010).

\bibitem{alfano2016optical}
R.~R. Alfano, G.~Milione, E.~J. Galvez, and L.~Shi,
  {\protect\JournalTitle{Nature Photonics}} \textbf{10}, 286 (2016).

\bibitem{sun2013large}
J.~Sun, E.~Timurdogan, A.~Yaacobi, E.~S. Hosseini, and M.~R. Watts,
  {\protect\JournalTitle{Nature}} \textbf{493}, 195 (2013).

\bibitem{shen2015integrated}
B.~Shen, P.~Wang, R.~Polson, and R.~Menon, {\protect\JournalTitle{Nature
  Photonics}} \textbf{9}, 378 (2015).

\bibitem{cai2012integrated}
X.~Cai, J.~Wang, M.~J. Strain, B.~Johnson-Morris, J.~Zhu, M.~Sorel, J.~L.
  O’Brien, M.~G. Thompson, and S.~Yu, {\protect\JournalTitle{Science}}
  \textbf{338}, 363 (2012).

\bibitem{lu2015broadband}
Z.~Lu, H.~Yun, Y.~Wang, Z.~Chen, F.~Zhang, N.~A. Jaeger, and L.~Chrostowski,
  {\protect\JournalTitle{Optics express}} \textbf{23}, 3795 (2015).

\bibitem{luo2014wdm}
L.-W. Luo, N.~Ophir, C.~P. Chen, L.~H. Gabrielli, C.~B. Poitras, K.~Bergmen,
  and M.~Lipson, {\protect\JournalTitle{Nature communications}} \textbf{5},
  3069 (2014).

\bibitem{dong2016128}
P.~Dong, X.~Chen, K.~Kim, S.~Chandrasekhar, Y.-K. Chen, and J.~H. Sinsky,
  {\protect\JournalTitle{Optics Express}} \textbf{24}, 14208 (2016).

\bibitem{taillaert2003compact}
D.~Taillaert, H.~Chong, P.~I. Borel, L.~H. Frandsen, R.~M. De~La~Rue, and
  R.~Baets, {\protect\JournalTitle{IEEE Photonics Technology Letters}}
  \textbf{15}, 1249 (2003).

\bibitem{tyo2002design}
J.~S. Tyo, {\protect\JournalTitle{Applied optics}} \textbf{41}, 619 (2002).

\bibitem{zhang2014high}
Y.~Zhang, S.~Yang, Y.~Yang, M.~Gould, N.~Ophir, A.~E.-J. Lim, G.-Q. Lo,
  P.~Magill, K.~Bergman, T.~Baehr-Jones \emph{et~al.},
  {\protect\JournalTitle{Optics express}} \textbf{22}, 11367 (2014).

\bibitem{derose2011ultra}
C.~T. DeRose, D.~C. Trotter, W.~A. Zortman, A.~L. Starbuck, M.~Fisher, M.~R.
  Watts, and P.~S. Davids, {\protect\JournalTitle{Optics express}} \textbf{19},
  24897 (2011).

\bibitem{seok2016large}
T.~J. Seok, N.~Quack, S.~Han, R.~S. Muller, and M.~C. Wu,
  {\protect\JournalTitle{Optica}} \textbf{3}, 64 (2016).

\bibitem{almeida2004all}
V.~R. Almeida, C.~A. Barrios, R.~R. Panepucci, M.~Lipson, M.~A. Foster, D.~G.
  Ouzounov, and A.~L. Gaeta, {\protect\JournalTitle{Optics Letters}}
  \textbf{29}, 2867 (2004).

\bibitem{moss2013new}
D.~J. Moss, R.~Morandotti, A.~L. Gaeta, and M.~Lipson,
  {\protect\JournalTitle{Nature photonics}} \textbf{7}, 597 (2013).

\bibitem{soref2010mid}
R.~Soref, {\protect\JournalTitle{Nature photonics}} \textbf{4}, 495 (2010).

\bibitem{souza2018fourier}
M.~C. Souza, A.~Grieco, N.~C. Frateschi, and Y.~Fainman,
  {\protect\JournalTitle{Nature communications}} \textbf{9}, 665 (2018).

\end{thebibliography}


\end{document}